\documentclass[]{aa}
\usepackage{natbib}
\usepackage{graphicx}
\usepackage{txfonts}
\usepackage[colorlinks=true, citecolor=blue]{hyperref}
\usepackage{lscape}
\usepackage{orcidlink}

\begin{document} 

\title{Near-infrared photometry of the central stars of planetary nebulae with the VVVX survey} 

        \author{
    Dante Minniti\inst{1,2}\thanks{To whom correspondence should be addressed; e-mail: vvvdante@gmail.com}, \orcidlink{0000-0002-7064-099X},
    Vasiliki Fragkou\inst{3} \orcidlink{0000-0002-8634-4204},
    Javier Alonso-Garc\'ia\inst{4}, 
   Daniel Majaess\inst{5}, 
     \and 
     Arianna Cortesi \inst{3,6}
        }

        \authorrunning{D. Minniti, et al.} 
        
\institute{
Instituto de Astrof\'isica, Depto. de. F\'isica y Astronom\'ia, Facultad de Ciencias Exactas, Universidad Andr\'es Bello, Av. Fern\'andez Concha 700, Las Condes, Santiago, Chile
        \and
Vatican Observatory, V00120 Vatican City State, Italy  
        \and
Observatório do Valongo, Universidade Federal do Rio de Janeiro, Ladeira do Pedro Antônio-43, Rio de Janeiro 20080-090, Brazil
               \and
Centro de Astronomía (CITEVA), Universidad de Antofagasta, Av. Angamos 601, Antofagasta, Chile                       
        \and
Mount Saint Vincent University, Halifax, Canada
        \and
Instituto de Física, Universidade Federal do Rio de Janeiro, 21941-972, Rio de Janeiro, RJ, Brazil}

        \date{Received ...; Accepted ...}
        \titlerunning{CSPN in the Near-IR}

        \abstract
       {Achieving accurate photometric characterizations of central stars of planetary nebulae (CSPNe) toward the galactic plane is significantly hindered by the high levels of interstellar extinction in these regions. However, near-infrared (NIR) observations offer a more effective alternative, as extinction is substantially reduced at these wavelengths.
        }
         {By mitigating the effects of interstellar extinction via NIR observations of the Galactic disk and bulge, we seek to improve the identification and characterization of CSPNe in these regions, enabling a deeper understanding of their properties and evolutionary status.
         }
         {
         We used NIR photometry from the VISTA Variables in the V\'ia L\'actea (VVV) survey and its extension VVVX to define the NIR photometry of a large sample of CSPNe recently identified with Gaia EDR3 data. We explored the optical and NIR properties of all CSPNe in our sample and  searched for eclipsing binary candidates among them by employing relevant catalogs. 
         }  
{
We present a homogeneous catalog of 1274 CSPNe, including their Z, Y, H, J, and K NIR magnitudes and errors. We also include our findings on the photometric properties of our sample. We report 14 CSPNe with a large IR excess indicating cool companions and/or surrounding discs and 56 eclipsing binary candidates.
  }
        { Based on the present VVVX CSPN catalog, we conclude that NIR photometry can prove valuable for  further and in-depth studies of CSPNe. Subsequent studies ought to focus on exploring the true nature of  the CSPNe that present IR excess as well as eclipsing-binary CSPNe candidates.
         }

        \maketitle
        
        \section{Introduction}
        \label{section1}

 Source catalogs generally provide large samples of objects that enable follow-up studies by the community.
 Galactic planetary nebulae (PNe) catalogs have been very useful for studies of this short phase of stellar evolution.
Some classic examples include the catalog compiled by Acker et al. (1992) as well as the more recent  Hong Kong/AAO/Strasbourg/H$\alpha$ (HASH) database by Parker et al. (2016). In this context, the identification of the central sources of PNe (CSPNe), whose study is essential for stellar evolution studies (e.g., Cummings et al. 2018; Marigo et al. 2020), is an additional challenging process, especially at low Galactic latitudes where there is substantial extinction and source confusion (e.g., Kerber et al. 2003).
 Indeed,  finding a PN does not guarantee the identification of its parent star, as the parent stars are faint objects cooling down rapidly.
In spite of the challenges, large numbers of sources started to become available. For example,
Kerber et al. (2008) cataloged 234 CSPNe, while Weidmann et al. (2020) presented an extended catalog of 620 CSPNe.
Subsequently, the advent of Gaia accurate astrometry enabled 
 Stanghellini et al. (2020) to find the parallaxes for 430 CSPNe, enabling the estimation of their masses. More recently, Gonzalez-Santamaria et al. (2021) published a large CSPN catalog containing 2035 CSPNe.
The CSPN catalog of Gonzalez-Santamaria et al. (2021) is the most complete collection published to date, with measured parameters based on Gaia EDR3 data, containing a small number of false positives in the cases of the true CSPN being beyond the reach of Gaia (Fragkou et al., 2025, ApJ submitted). This catalog more than doubles the previous samples (e.g., Frey et al. 2013, Gonzalez-Santamaria et al. 2019, Weidmann et al. 2020, Chornay et al. 2021) and it also contains  a thorough description of the difficulty in identifying  central stars,  extinction corrections, distance measurements,  binarity, and  properties of the population of CSPNe as a whole. The extinction values that used in the present work were also obtained from the literature.

Importantly, the identification of large samples of CSPNe enables the search for binaries in this population, which is highly relevant as binarity is probably playing a role in shaping the morphology of PNe themselves (De Marco 2009). Furthermore, eclipsing binaries in particular may also help to measure dynamical masses.
For example, Barker et al. (2018) discussed the potential of the VPHAS+ survey to detect binary CSPNe.
Chornay \& Walton (2020)  published a catalog of CSPNe identified using Gaia DR2 data, a handful of which were later confirmed as binaries by Chornay et al. (2021). Bhatthacharjee et al. (2025) used the Zwicky Transient Factory (ZTF) to discover 94 variable CSPNe, nearly doubling the previously known sample.
In addition, Kepler and TESS have provided large variability databases with exquisite photometric scatter, allowing for significant progress to be made in this area thanks to their confirmation that the binary fraction of CSPNe is high.
Jacobi et al. (2021) studied 204 CSPNe candidates identified with the Kepler/K2 mission, 36 of which turn out to be periodic light curve variations, usually associated with eclipsing binaries. 
Aller et al. (2024) studied 38 short-period eclipsing binaries observed with TESS, discovering that 24 in total showed periodic variations.

There are  a number of studies about the binary fraction of CSPNe (see Barker et al. 2018; De Marco et al. 2013; the review by Jones 2020 and references therein).
Miszalski et al. (2009a, b) discovered 21 eclipsing CSPNe binaries using the OGLE survey data, inferring a significant close binary fraction among CSPNe, many of which appear to be bipolar nebulae. 
However, the binary fraction measured from space by Jacobi et al. (2021), which is a lower limit of 20–24 $\pm$ 5  \%, is significantly larger than previous ground-based estimates (13 \% from Bond 2000, and 12–21 \% from Miszalski et al. 2009).

Binary CSPNe are important laboratories to study the common envelope phase of binary evolution.
For instance, authors such as De Marco et al. (2009), Jones \& Boffin (2017; 2019), Hillwig et al. (2017), Decin et al. (2020) have thoroughly discussed how asymmetric PNe shapes are related to binary companion interaction, as the formation and evolution of these objects would be different from the PNe associated with isolated single stars. Jones (2020) offers a summary of the current understanding of the role of binarity in the PN formation and evolution during the common envelope phase.

Jiménez-Esteban et al. (2019) searched for wide binaries among CSPNe using Gaia DR2 astrometric parameters to detect comoving objects. More recently, Gonzalez-Santamaria et al. (2021) applied the same technique to search for resolved wide astrometric binaries in their CSPN catalog, while also carrying out a statistical procedure in order to detect possible close binaries based on Gaia EDR3 data. The latter technique is based on the assumption that  statistically noisy Gaia measurements (e.g., astrometric excess, renormalized unit weight error, harmonic amplitudes, and coordinate uncertainties, or anomalous red colors) are an indication of a close binary companion.
Chornay et al. (2021) also searched for variable (i.e., eclipsing binary) CSPNe using statistical fluctuations in the Gaia DR2 photometry, confirming four genuinely variable stars consistent with binarity.

In particular, Gonzalez-Santamaria et al. (2021) concluded that there was evidence for the red CSPNe (with dereddened colors BP-RP>0.2 mag) to be statistically different from the blue sample of CSPNe (with dereddened colors BP-RP<0.2 mag), indicating a greater incidence of close binarity among the red CSPNe population. However, in the statistical case the assumptions need to be confirmed; in addition, projection effects cannot be discarded as opposed to real physical close binaries. In this context, it is important to search for eclipsing binary CSPNe in their sample to confirm the statistical conclusions.

This paper presents the VVVX near-infrared (NIR) photometry catalog containing 1274 CSPNe based on the match of our database with the CSPN catalog of Gonzalez-Santamaria et al. (2021). Section 2 contains a description of our survey and previous work on PNe.
Section 3 presents the CSPN catalog, discussing the most compact PNe, the cases of high extinction and NIR excess, along with  potential misidentifications.
Section 4 contains the selected sample of binary CSPNe, including eclipsing binaries.
Finally, our conclusions are summarized in Section 5.
A follow-up paper will discuss the PNe associated with star clusters (Fragkou et al., in prep.).

\section{VVV NIR data and previous work}
        \label{section2}

A series of NIR public surveys have systematically mapped the Southern Galactic plane and bulge using VIRCAM at the VISTA 4m telescope at ESO Paranal Observatory in Chile (Minniti et al. 2010, Saito et al. 2024). The ESO public survey VISTA Variables in the Vía Láctea (VVV) surveyed the inner Galactic bulge and the adjacent southern Galactic disk in the Z, Y, J, H, and Ks NIR filters for 2000 hours of observations in total  spanning from years 2009 to 2015 (Minniti et al. 2010, Saito et al. 2012). 
Upon its conclusion, the complementary VVV extended (VVVX) survey expanded both the temporal and spatial coverage of the original VVV area, widening it from 562 to 1700 sq. deg., while providing additional epochs in
the J, H, and Ks  filters from 2016 to 2023  (Saito et al. 2024). 
The VVVX survey took about 2200 hours of observation to complete, covering about 4\% of the sky in the bulge and southern Galactic disk. 
VVVX covered most of the gaps left between the VVV and the VISTA Hemisphere Survey (VHS) areas and extended the VVV time baseline in the obscured regions affected by high extinction (and hence hidden from optical observations). 
The VIRCAM instrument on board the VISTA telescope provided $1.5 \times 1.1$ sq.deg. wide-field images with a resolution of 0.34"  per pixel.
From those images, we identified sources and extracted their photometries using point spread function (PSF)-fitting techniques (Alonso-García et al. 2018), generating catalogs typically reaching a limiting magnitude of $Ks = 18$ mag throughout the Galactic plane, which is reduced to approximately  $Ks = 17$ mag within the inner Galactic bulge because of crowding.

These NIR survey data have served  a number of studies from the interstellar medium to variable stars, from the search for star clusters and exoplanets to background galaxies and QSOs (e.g., Alonso-García et al 2025).
Within the existing VVV area, we produced a 5D map of the surveyed region by combining positions, distances, and proper motions of star-forming regions in the Galactic disk, as well as known distance indicators such as red clump stars, RR Lyrae, and Cepheid variables. 
In addition, the VVV+VVVX catalogs complement those from the Gaia mission at low Galactic latitudes and provide spectroscopic targets for the forthcoming  high-multiplex spectrographs such as MOONS, 4MOST, and WST.
Some of our previous works on PNe with the VVV that are relevant to this study include 
Weidmann et al. (2013), presenting initial results for NIR photometry of 353 known Galactic PNe with the VVV Survey, and 
Minniti et al. (2019), who presented new candidate PNe associated with Galactic globular clusters.
In addition, some of us have also been involved in studies of PNe associated
with Galactic open clusters (e.g., Fragkou et al. 2019a,b, 2022, 2025 and
Majaess et al. 2007).

\section{The VVV/VVVX CSPN catalog}
        \label{section3}

To build the CSPN catalog we crossmatched our NIR VVV/VVVX catalogs with the sample of CSPNe from Gonzalez-Santamaria et al. (2021), which is the largest sample available. 
We selected a search radius of $0.7$" in order to make the match between the VVV+VVVX NIR photometry with the Gaia optical photometry from the CSPNe of Gonzalez-Santamaria et al. (2021). However, the distribution of final separations reveal that the positions of both samples are so accurate that the coincidence is always better than 0.6".             
There are 1274 matches in total (see Table A1), including 
$N_A= 343$ (27 \%) that belong to their group A (sources with high reliability of being CSPNe), 
$N_B= 420$ (33 \%) that belong to their group B (sources with intermediate reliability of being CSPNe), and 
$N_C= 511$ (40 \%) that belong to their group C (sources with low reliability of being CSPNe). The assigned reliability of each source (high, intermediate or low) is based on both its Gaia $BP-RP$ dereddened color (very blue stars are more reliable CSPN candidates) and separation from the apparent center of its respective PN (for a more detailed discription of the assignment of the groups see Gonzalez-Santamaria et al. 2021). The coordinates listed in all tables correspond to the CSPNe. We matched separately the $ZYJHKs$ photometry from the VVV survey and the $JHKs$ photometry from the VVVX survey to keep track of these different samples. 

A total of 510 of our matches have a classification in SIMBAD (Wenger et al. 2020), mostly as PNe, white dwarfs, and AGB stars as expected. Twelve of all matches are flagged as eclipsing binaries. Figure 1 shows the map of the Gonzalez-Santamaria et al. (2021) CSPNe matched with the VVV survey (red circles) and the VVVX survey (blue circles). 
The higher spatial concentration of the CSPNe to the Galactic bulge is evident, while the disk region exhibits a more uniform and less dense distribution. 
A sample of the data for the present CSPN catalog are shown in Table A1, where we list the PN identifications, positions RA and DEC (J2000), optical magnitudes G and $BP-RP$ colors, NIR magnitudes $ZYJHKs$, and the group ABC for some of individual sources (the full catalog is available online). 

        \begin{figure*}
        \begin{center}
                \includegraphics[width=180mm]{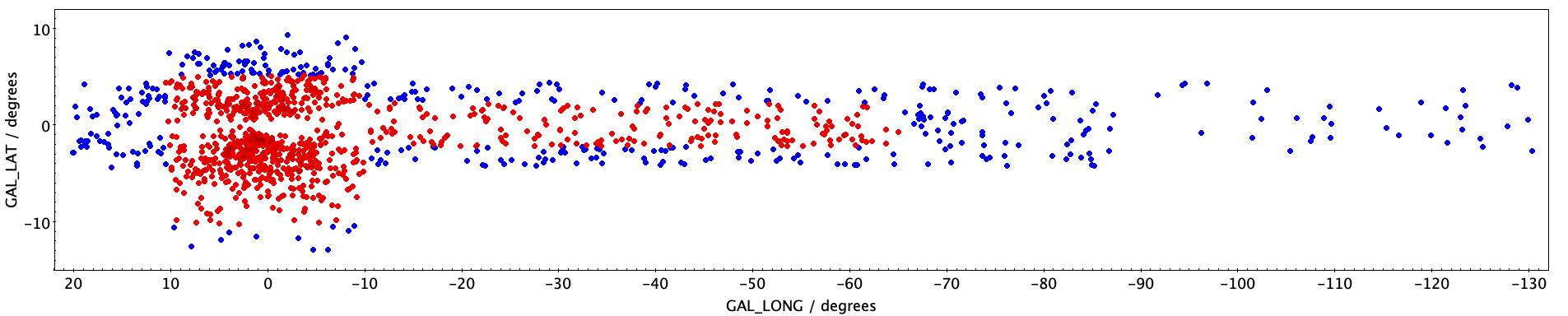}
                \caption{
Map of the Gonzalez-Santamaria et al. (2021) CSPNe matched with the VVV survey (red circles) and with the VVVX survey (blue circles). 
              }
                \label{Figure2}
        \end{center}
\end{figure*}

Figure 2 shows the comparison of the CMDs for the VVV versus VVVX samples matched CSPNe from Gonzalez-Santamaria et al. (2021).
We observe that the optical and NIR CMDs show wide distributions, with no large systematic differences between the two surveys; however, we note that the VVVX sample, generally located away from the crowded and obscured inner bulge regions, tends to be bluer. 
The same is true for the optical-NIR color-color diagram shown in the right panel of Figure 2.

        \begin{figure*}
        \begin{center}
                \includegraphics[width=\textwidth]
                {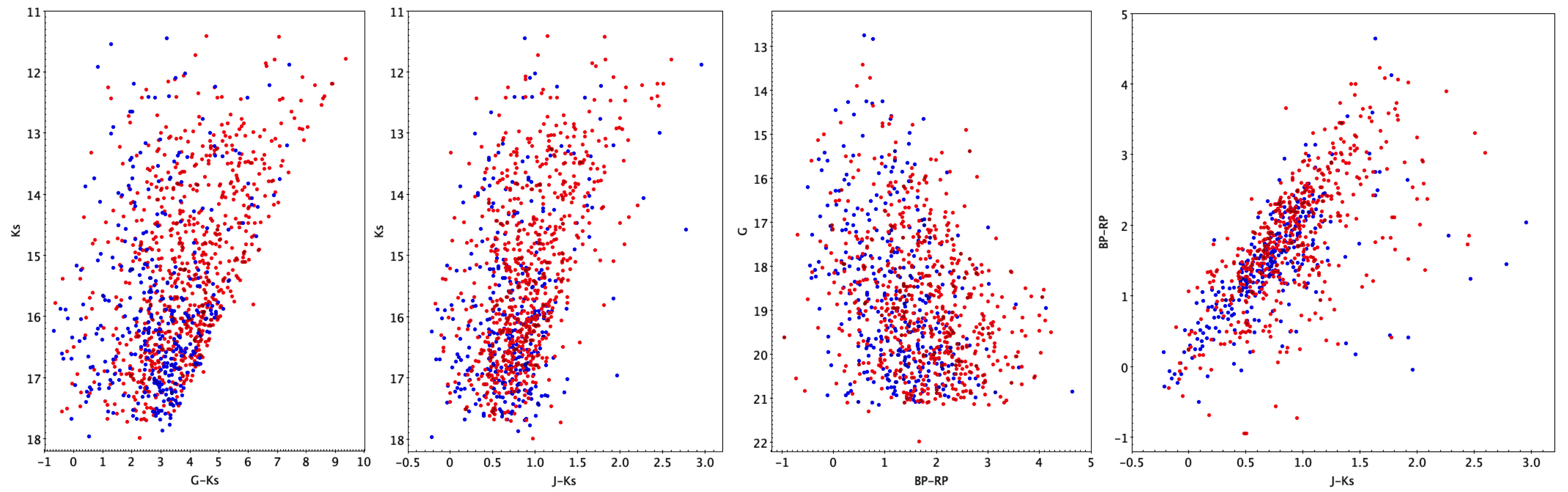}
                \caption{
Comparison of the VVV vs VVVX samples matched CSPNe from Gonzalez-Santamaria et al. (2021), displaying the optical and NIR CMDs and color-color diagram for the VVV survey (red circles) and the VVVX survey (blue circles). 
                }
                \label{Figure3}
        \end{center}
\end{figure*}

It is also important to check which objects are present in the Gonzalez-Santamaria et al. (2021) CSPN catalog, but absent in our NIR data.
We can use the optical CMD based on Gaia photometry to make this comparison.
Figure 3 shows the CSPNe of Gonzalez-Santamaria et al. (2021) matched with the VVV and VVVX survey (red circles), compared with the unmatched original sources (grey crosses). 
This optical CMD shows a similar overall distribution, with no large differences between these two samples. 
We argue that we did miss some of the brightest objects, which  were not matched because of saturation in our NIR images; in addition,  some of the faintest and bluest sources were unmatched because they lay beyond our NIR detection limit ($Ks \sim 17-18$ mag). 
Likewise, there are N=238 CSPNe from Gonzalez-Santamaria et al. (2021) that do not have Gaia colors and are therefore not plotted in this CMD.

        \begin{figure}
        \begin{center}
               \includegraphics[width=60mm]{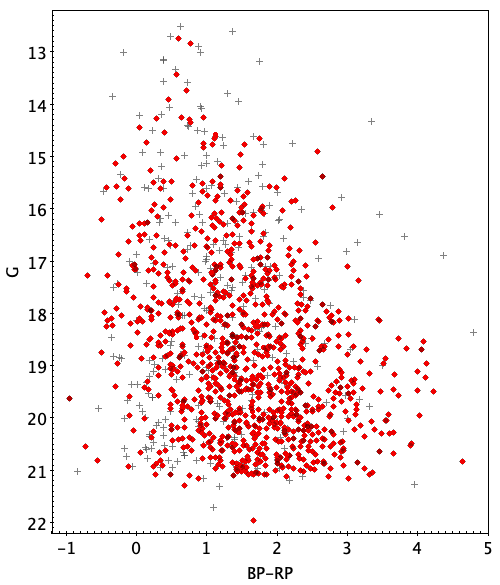}
                \caption{
{
CSPNe of Gonzalez-Santamaria et al. (2021) matched with the VVV and VVVX survey (red circles), compared with the unmatched original sources (grey crosses).                             }
}
                \label{Figure4}
        \end{center}
\end{figure}

Gonzalez-Santamaria et al. (2021) used the extinction maps of Schlegel et al. (1998) and Green et al. (2019) to estimate the $A_V$ for the individual sources. 
Figure 4 shows the magnitude dependence on extinction and optical color distribution of the CSPN samples.
In general, the observed colors for the CSPNe belonging to their group C (more uncertain candidates) are redder than those of the group AB. 
However, the $A_V$ optical extinction distributions of group AB and group C seem to be similar to each other. 
So even after correcting for reddening, the mean color of group C CSPNe is redder than group AB. 
It is interesting to confirm that this is true also in the NIR to discard any kind of selection effect with color.

        \begin{figure}
        \begin{center}
                \includegraphics[width=90mm]{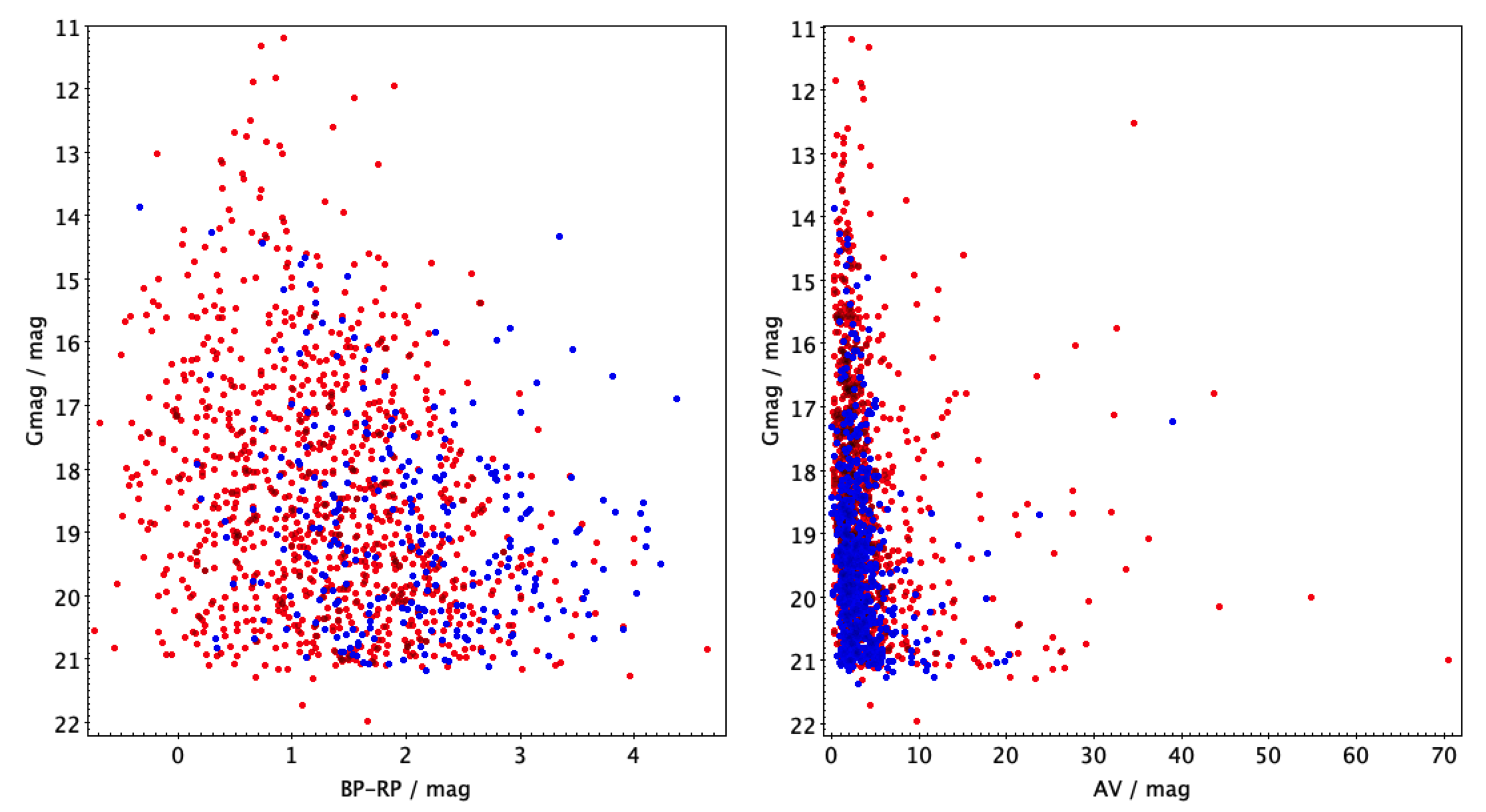}
                \caption{
{
Optical CMD and magnitude dependence on extinction distribution of the CSPN AB (red) and C (blue) subsamples.
                                  }
}
                \label{Figure5}
        \end{center}
\end{figure}

Figure 5 shows the comparison with all matched Gonzalez-Santamaria et al. (2021) CSPNe, including group AB (red) versus group C (magenta).
We confirm that even in the NIR, the general trends for the observed colors for the CSPNe of group C of Gonzalez-Santamaria et al. (2021) (more uncertain candidates) are redder than those of the group AB. 
There are $N=238$ objects in the group C sample of CSPNe from Gonzalez-Santamaria et al. (2021) that do not have Gaia colors, which they assigned to BP-RP=0. 
Their NIR colors are normal, suggesting that  the majority of them had been correctly identified as CSPNe.

        \begin{figure*}
        \begin{center}
                \includegraphics[width=\textwidth]{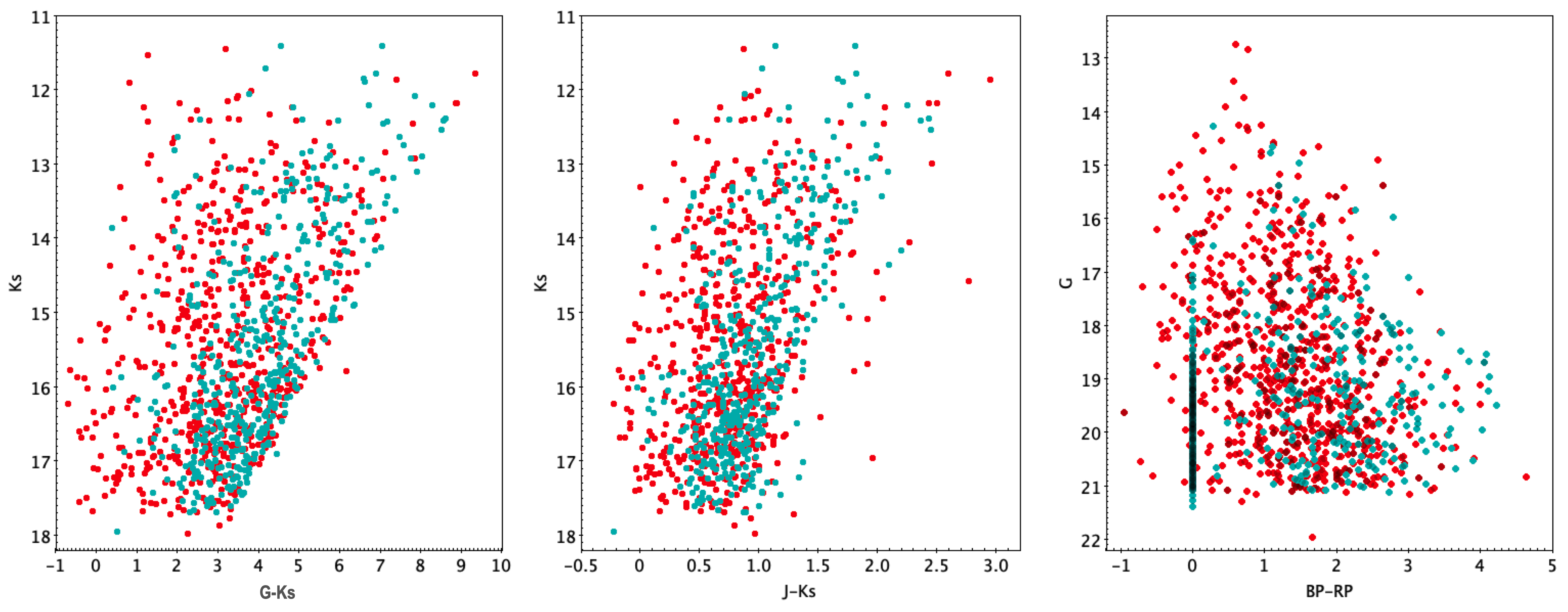}
                
                \caption{
{
Comparison with all matched Gonzalez-Santamaria et al. (2021) CSPNe, including the group AB (red) vs group C (magenta). Note: all sources without Gaia optical colors have been assigned to $BP-RP=0.0$ mag.
                                  }
}
                \label{Figure6}
        \end{center}
\end{figure*}

Figure 6 shows the optical and NIR color distributions for all matched Gonzalez-Santamaria et al. (2021) CSPNe with the stars without BP-RP colors.
We note that a significant fraction of CSPNe from Gonzalez-Santamaria et al. (2021) are too faint to possess a measurable Gaia BP-RP color; whereas our VVV+VVVX surveys provide for the first time the only colors available to complement the data for these objects. 
These $N=238$ objects (17 \%) without BP-RP color are shown in blue. Gonzalez-Santamaria et al. (2021) assign $BP-RP=0$ to these objects without Gaia colors. 
Except for the fact that these $N=238$ tend to be fainter in the optical, there are no systematic differences with the rest of the whole population in $J-Ks$, $G-Ks$, $H-Ks$, and $Z-Y$.

        \begin{figure}
        \begin{center}
                \includegraphics[width=90mm]{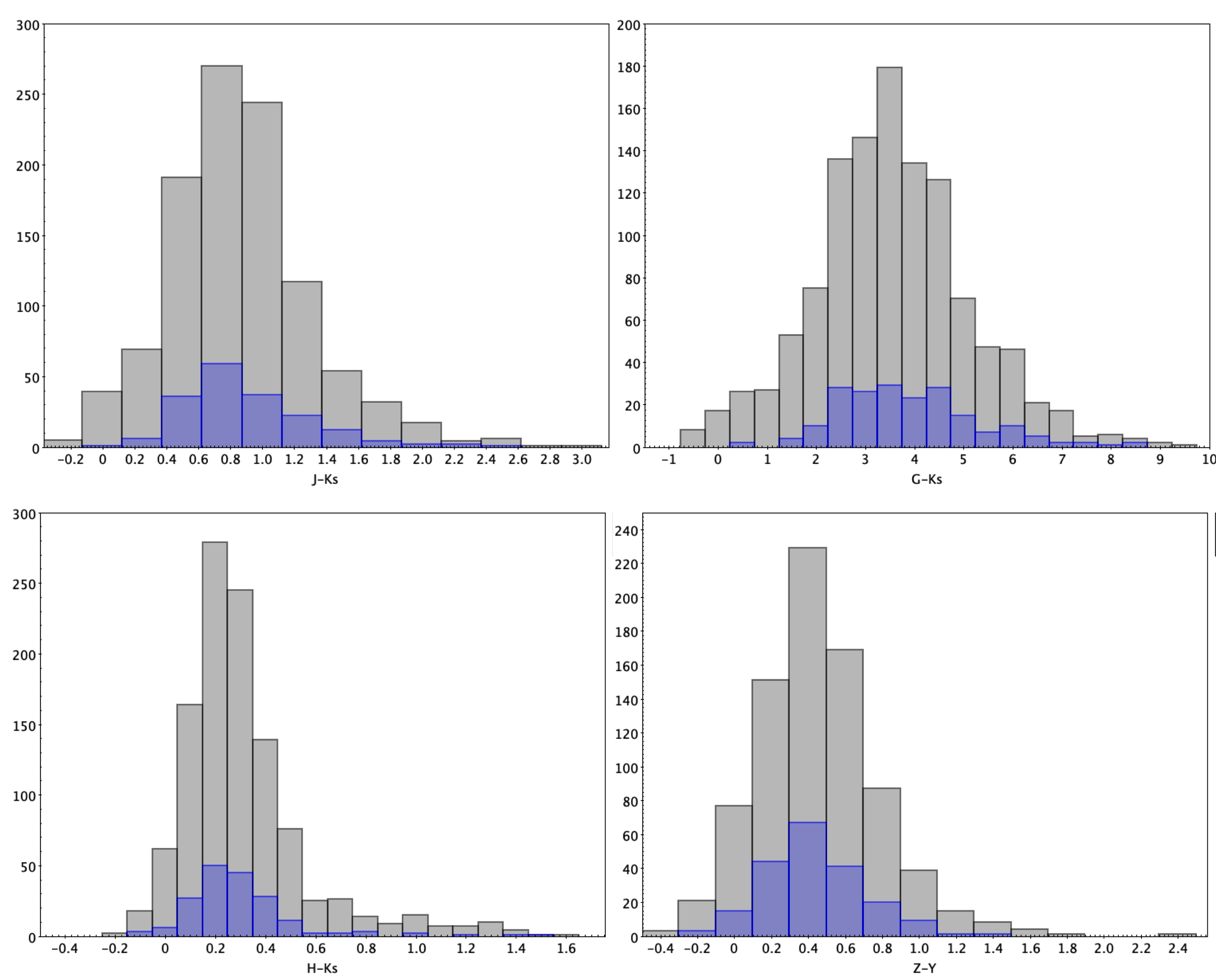}
                \caption{
{
 $J-Ks$, $G-Ks$, $H-Ks$, and $Z-Y$ color distributions comparison with all matched Gonzalez-Santamaria et al. (2021) CSPNe, including stars without BP-RP colors (blue histograms).
 These histograms show no significant differences between the samples.
                }
}
                \label{Figure7}
        \end{center}
\end{figure}

\subsection{The CSPNe for the most compact PNe}

We also inspected the CSPNe for the most compact PNe, selecting the objects with angular major diameters $dang \leq 5$ arcsec, as listed in the HASH database (Parker et al. 2016). 
These are expected to primarily be very distant objects, along with some PNe in the early stages of evolution.
From the total of 1240 objects in our sample with measured angular major diameters, there are 109 such compact objects, representing roughly 9 \% of the total sample.
In spite of the fact that many of them lack the optical colors, we notice that there are no trends with magnitude or colors in the optical and NIR CMDs (see Figure 7), indicating that there is no significant bias towards a more reddened or a more distant population. 

        \begin{figure}
        \begin{center}
                \includegraphics[width=\linewidth]{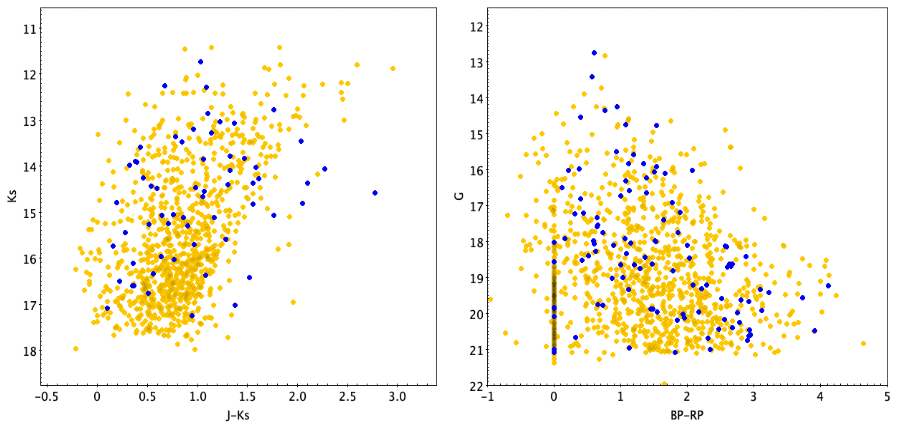}
                \caption{
{
Comparison with the CSPNe of the most compact PNe  ($dang \leq 5$ arcsec) shown in blue.
Note: a large number of compact objects do not have Gaia optical  colors, which have been arbitrarily assigned to $BP-RP=0.0$ mag.
                                             }
}
                \label{Figure8}
        \end{center}
\end{figure}

\subsection{High extinction and infrared excess}

The optical extinction $A_V$ versus different colors exhibit a lot of scatter, but the majority of the CSPNe are distributed in a well defined wide sequence, 
where the general trend is for increasing optical and NIR colors with increasing $A_V$, as expected (see Figure 8). 
Then there are several objects to the top of the diagram that exhibit very high extinctions. 
These include 90 objects with $A_V >10$ mag (representing 7~\% of the sample of $N=1274$ objects). 
Many of these objects do not have redder NIR colors, indicating that their $A_V$ extinctions may be flawed.
Also we noticed the presence of a few low extinction objects with large color excess in these diagrams. 
The fact that they have excess in different color combinations argues for a real observational effect.

        \begin{figure*}
        \begin{center}
                \includegraphics[width=\linewidth]{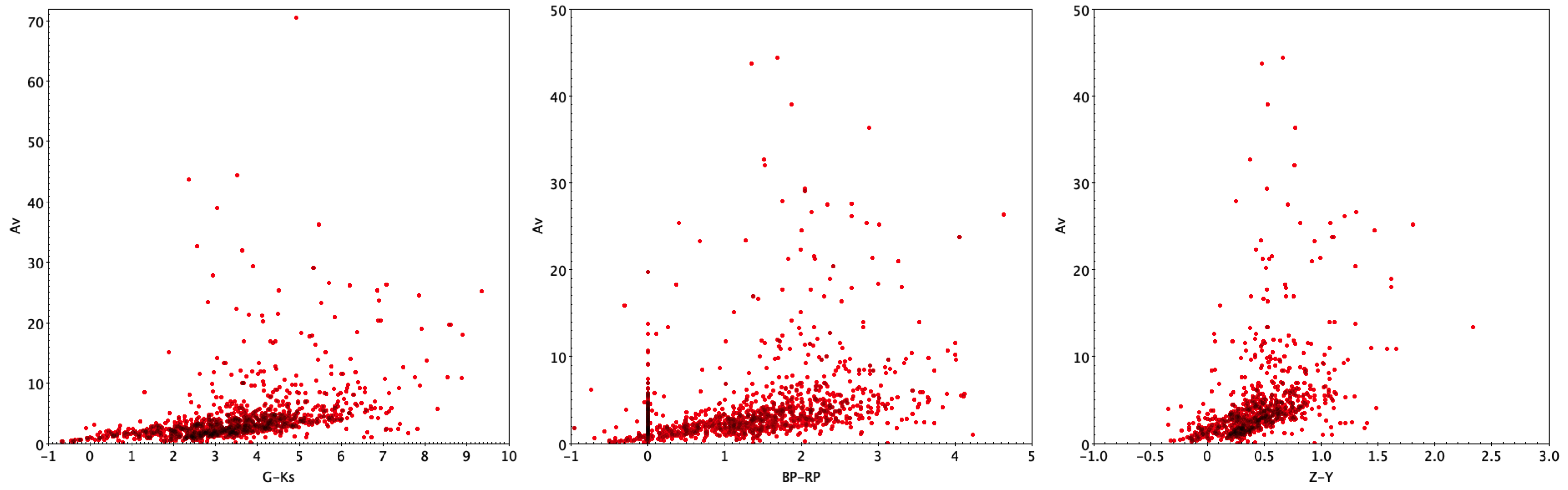}
                \caption{
Optical and NIR color trends as function of the optical extinction $A_V$ assigned by Gonzalez-Santamaria et al. (2021). Stars without existing Gaia color measurements have been arbitrarily assigned $BP-RP=0.0$ mag.
}
                \label{Figure9}
        \end{center}
\end{figure*}

Investigating  the objects with very high extinctions, we selected the sources listed by Gonzalez-Santamaria et al (2021) to have  very high extinction, arbitrarily chosing $A_V>10$ mag (equivalent to $A_K>1.1$ mag). 
These are highlighted with large red circles in the CMDs and CCD shown in Figure 9.
As expected, these objects with $A_V>10$ mag have on average redder NIR colors, even though they do not seem to form a well defined group separate from the general population.

        \begin{figure*}
        \begin{center}
                \includegraphics[width=\linewidth]{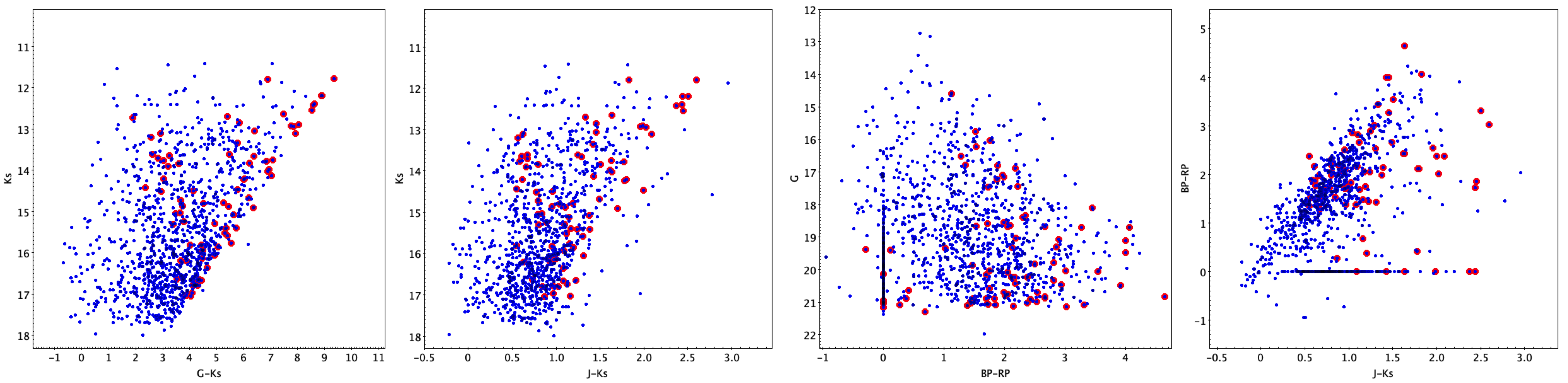}
                \caption{
Optical and NIR CMDs and color-color diagram showing the selection of CSPNe with very high measured optical extinction $A_V>10$ mag (red circles) . Again, the sources without Gaia color measurements have been arbitrarily assigned $BP-RP=0.0$ mag.
}
                \label{Figure10}
        \end{center}
\end{figure*}

A large IR color may not only be due to high reddening, but also due to the presence of a cool companion (e.g., De Marco et al. 2013, Douchin et al. 2015, Barker et al. 2018). The NIR excess can be used to search for cool companion stars that may be orbiting around the central PN ionizing hot star (e.g., Douchin et al. 2015), provided that the nebular emission is taken into account. However, in the Galactic plane and bulge, this method is hampered by the large and non-uniform reddening and extinction across these Galactic regions.
Alternatively, in the cases where the NIR excess is very large, instead of the presence of a cool companion it may indicate the presence of circumstellar disks. 
It is therefore important to try to discriminate between these different possibilities. Nebular internal extinction can be significant for some young compact PNe and affect the observed IR excess, but since it is negligible for most PNe (Frew et al. 2016), it is not expected to significantly affect our results.
Figure 10 shows the selection of CSPNe with IR excess, namely, sources that are too red for their individual extinction.
 In the $A_V$ $vs$ $G-Ks$ diagram there are a few objects that have relatively low optical extinctions ($Av<3.3$ mag, $E(B-V)<10$ mag), but very large colors, indicative of NIR excess. 
We selected  objects located in the bottom right of the diagram,  presenting a $G-Ks$ color 1 to 3 mag larger than the $G-Ks$ color of the general distribution, which lies between $G-Ks=$ 2 - 4.5 mag and $A_V=$ 0.2 - 4.5 (i.e., this is indicative of NIR excess).
 There are $N=14$ objects selected (see Table 1), that are prime candidates to contain cool companions and/or disks surrounding the CSPN.    

         \begin{figure}
        \begin{center}
                \includegraphics[width=\linewidth]{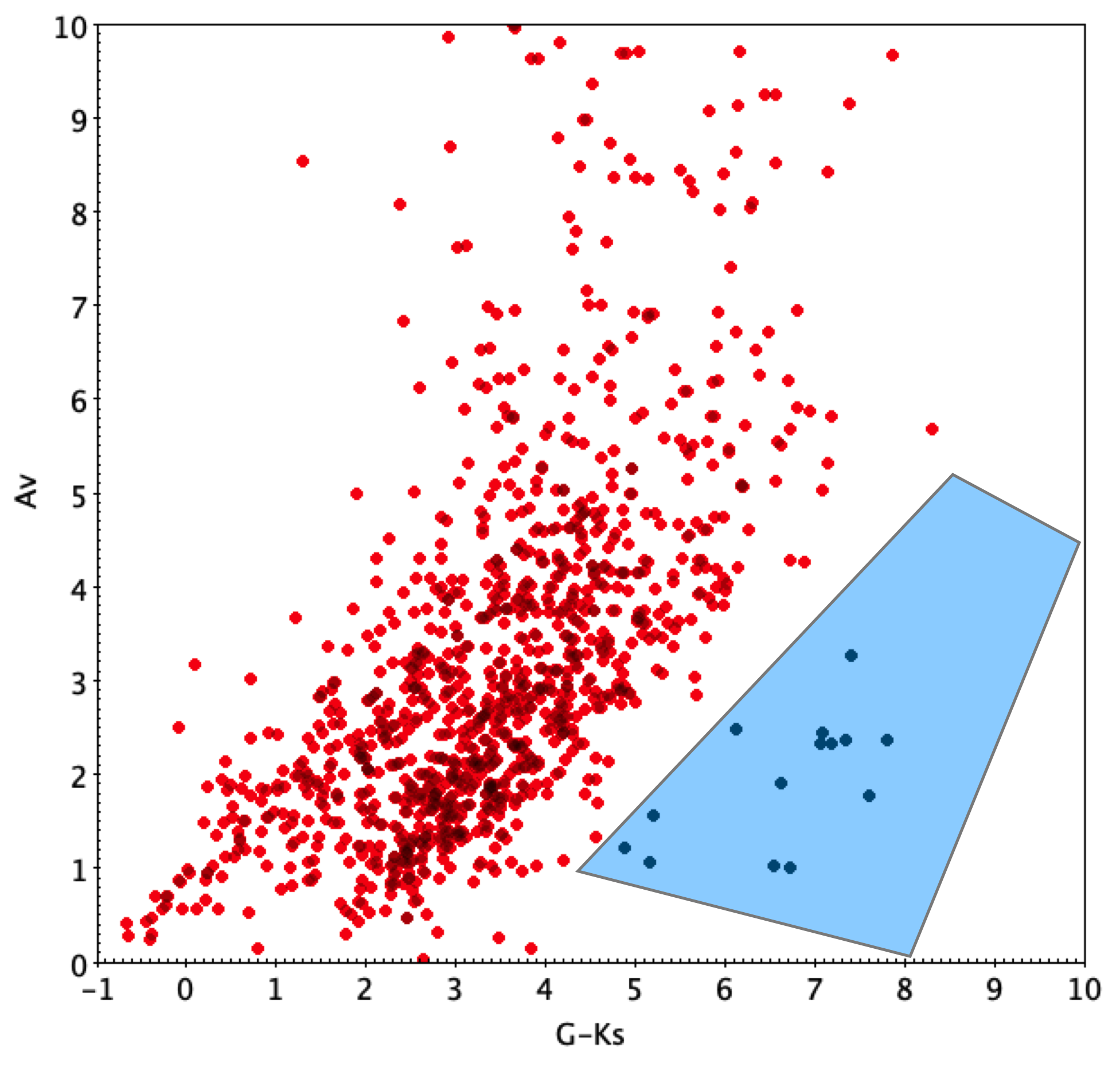}
                \caption{
Optical extinction vs color illustrating the selection of objects that exhibit overly high IR excess for their assigned optical extinction (light blue region).
}
                \label{Figure11}
        \end{center}
\end{figure}

\begin{table}[]
\caption{CSPNe that present IR excess.}
\label{tab:Tab1}
\begin{tabular}{llcccc}
\hline \hline
RA (deg) & DEC (deg) & G mag & Ks mag & $A_V$   & Group \\
\hline
267.4617  & -30.0530   & 20.64  & 13.45   & 2.32 & C     \\
267.5433  & -29.3622   & 18.47  & 11.85   & 1.90 & C     \\
267.8533  & -28.5948   & 20.31  & 13.58   & 1.01 & C     \\
268.1907  & -28.8108   & 20.10  & 15.21   & 1.22 & C     \\
268.3970  & -28.4808   & 18.16  & 12.99   & 1.06 & B     \\
268.7933  & -27.6944   & 19.50  & 12.95   & 1.02 & C     \\
268.9830  & -22.9837   & 20.30  & 13.21   & 2.45 & B     \\
269.1607  & -26.1383   & 20.36  & 12.76   & 1.77 & C     \\
269.7709  & -20.4570   & 18.49  & 11.42   & 2.33 & C     \\
270.1764  & -26.8938   & 19.68  & 14.48   & 1.57 & C     \\
270.1800  & -17.7553   & 19.28  & 11.88   & 3.26 & B     \\
271.3118  & -23.7859   & 20.28  & 12.47   & 2.37 & B     \\
276.8065  & -14.1430   & 20.55  & 13.20   & 2.37 & C     \\
277.6775  & -13.5183   & 20.30  & 14.18   & 2.48 & C   \\
\hline
\end{tabular}
\tablefoot{
CSPNe from Gonzalez-Santamaria et al. (2021) that present IR excess based on their Gaia and VVV+VVVX magnitudes.}

\end{table}

There are only 14 objects with such a high IR excess, but this number should be taken as a lower limit to the frequency of this population, because  this CSPN sample of Gonzalez-Santamaria et al. (2021) was selected using optical photometry from Gaia.
 As expected, the selected sample of 14 objects with IR excess occupy the red part of the optical and NIR CMDs and color-color diagram (four of them belonging to group B and eight to group C). 
 They are bright in the NIR, but very faint and red in the optical Gaia CMDs (Figure 11). 

         \begin{figure*}
        \begin{center}
                \includegraphics[width=\linewidth]{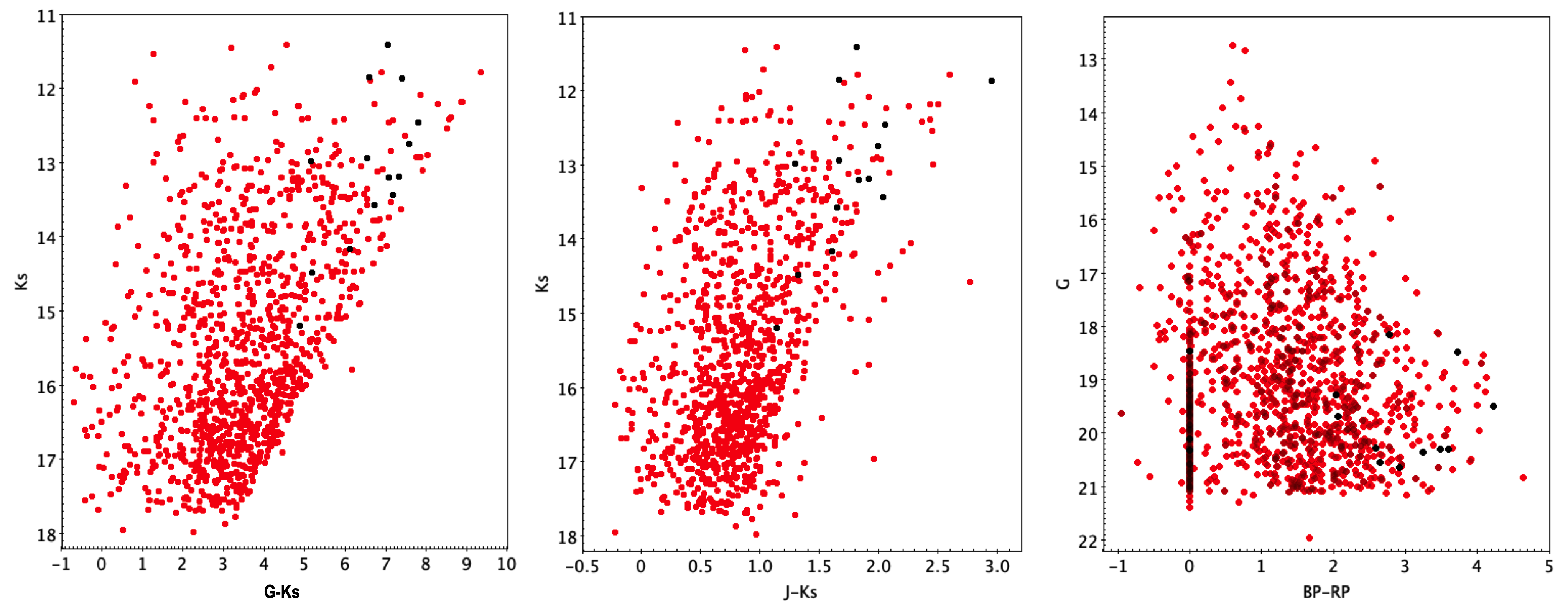}
               
                \caption{
Optical-NIR CMDs showing the IR excess objects highlighted in black.
Note: a large number of compact objects do not have Gaia optical $BP-RP$ colors, which have been arbitrarily assigned $BP-RP=0.0$ mag.
}
                \label{Figure12}
        \end{center}
\end{figure*}

The optical-NIR CCD for the whole sample shows an interesting distribution (Figure 12).
There appears to be a separate sequence of unreddened CSPNe.  There is also a scattered subsample that has redder J-Ks colors compared with the BP-RP colors. These extreme IR colors are much more intense than can be accounted for with the photometric scatter; thus, they are likely to be revealing IR excess due to the presence of a cool companion.
Therefore, we single out these IR-excess objects  as interesting candidates with respect to  their having cool companions whose nature must be verified with additional data.

        \begin{figure}
        \begin{center}
                \includegraphics[width=\linewidth]{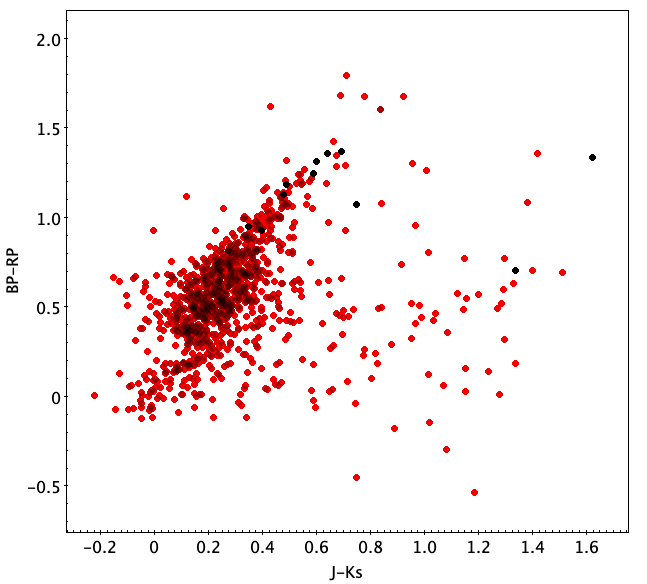}
                \caption{
Optical-NIR color-color diagram for the CSPN sample,  with the IR excess objects again highlighted in black, as in the previous figure.
}
                \label{Figure13}
        \end{center}
\end{figure}

\subsection{Misclassifications}

We  also conducted a brief check for potential misclassifications. 
Inspecting the galaxy catalog of Baravalle et al. (2023), we find only one object in common that may have been misclassified as a galaxy. 
This source is PNG305.9-01.6 = MPAJ1319-6418, located at Galactic coordinates $GL=305.95683$ deg, $GB= -1.60745$ deg within the VVV survey footprint.
The Gaia photometry gives: $G=20.97$ mag, $BP-RP=1.13$ mag,
while the VVV photometry for this source yields: $Z=19.83 \pm 0.12$ mag, $Y=18.35 \pm 0.08$ mag, $J=16.84 \pm 0.03$ mag, $H=16.36 \pm 0.03 $ mag, and $Ks=14.94 \pm 0.02$ mag.
This object is also unusual because it is located within a few arcseconds of a very compact dark globule (likely in the foreground), as shown by an inspection of the DECAPS and VVV images (Figure 13).                
We conclude that PNG305.9-01.6 is a  CSPN because it looks like a point source and also because the colors are too blue to be a background galaxy seen throughout the Milky Way disk.
Moreover, eight of our objects are identified as candidate young stellar objects in SIMBAD but more data are needed to clarify their nature, which is beyond the scope of this study.

        \begin{figure}
        \begin{center}
                \includegraphics[width=\linewidth]{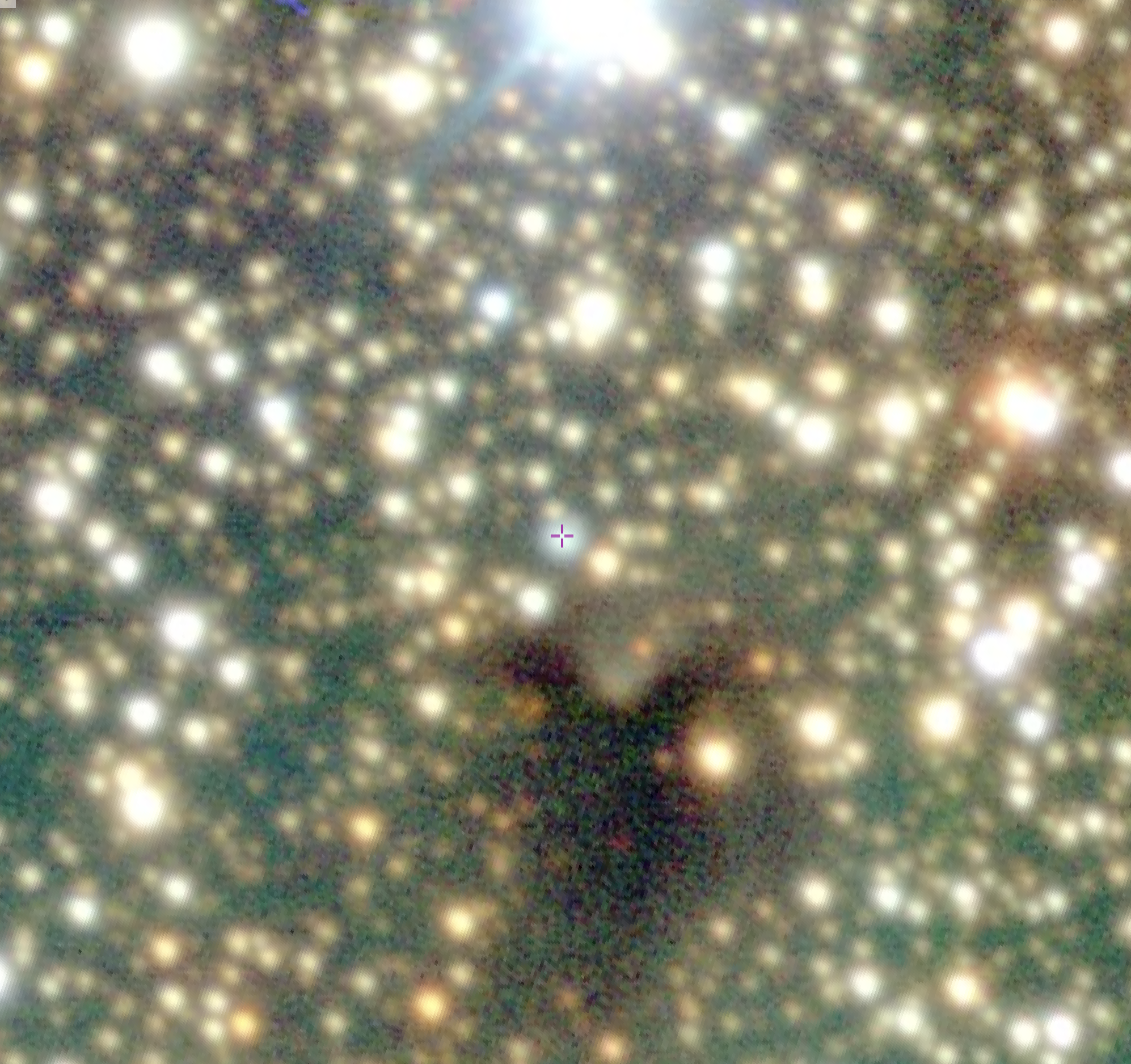}
                \caption{
 Field centered on PNG305.9-01.6 = MPAJ1319-6418, the only object that may have been misclassified as a galaxy (red plus sign). 
This optical color image is from the DECAPS survey (Schlafly et al. 2018), covering $90" \times 90"$, 
oriented along Galactic coordinates, with GL increasing to the left and GB increasing to the top.               
}
                \label{Figure14}
        \end{center}
\end{figure}

\section{Search for eclipsing binary CSPNe}
        \label{section5}

As an example of how our catalog can be applied, we searched for eclipsing binaries CSPN given these objects would be prime candidates for direct spectroscopic mass determinations. 
This search is complementary to the work of Gonzalez-Santamaria et al. (2021), which  only searched for wide binaries.
They found a number of wide binaries in the Gaia database with projected separations below 20000 AU. They also found statistical evidence to support the presence of unresolved pairs using some of the parameters measured by Gaia. 

The search for binaries is done in two ways here.
First, to search for variable CSPNe we can use the Virac Variable Classification Ensemble (VIVACE), a large catalog of NIR variables within the VVV footprint made by Molnar et al. (2022). 
We note that this is limited to the VVV survey because no such massive variable star catalog exists for the extended area of the VVVX survey yet.
Figure 14 shows the CMD resulting of the match between the samples of  
Gonzalez-Santamaria et al. (2021) in the optical and the VIVACE sample of NIR variables from Molnar et al. (2022).
We selected the stars classified as binaries in the VIVACE catalog, revealing
only 25 CSPN eclipsing binaries in common, with
only 12 best candidates (because either their magnitudes are too bright, or their amplitudes or periods too large).
Figure 15 shows the magnitude and amplitude dependence on period for this selected binary sample.
The mismatches and incompleteness in the eclipsing binary sample may also be due to the uneven Gaia observational footprint. Only five of these CSPN eclipsing binaries have well determined distances
from Gonzalez-Santamaria et al. (2021).
These objects are located relatively nearby, with $1.3 < Distance < 2.2$
kpc.

        \begin{figure*}
        \begin{center}
                \includegraphics[width=\linewidth]{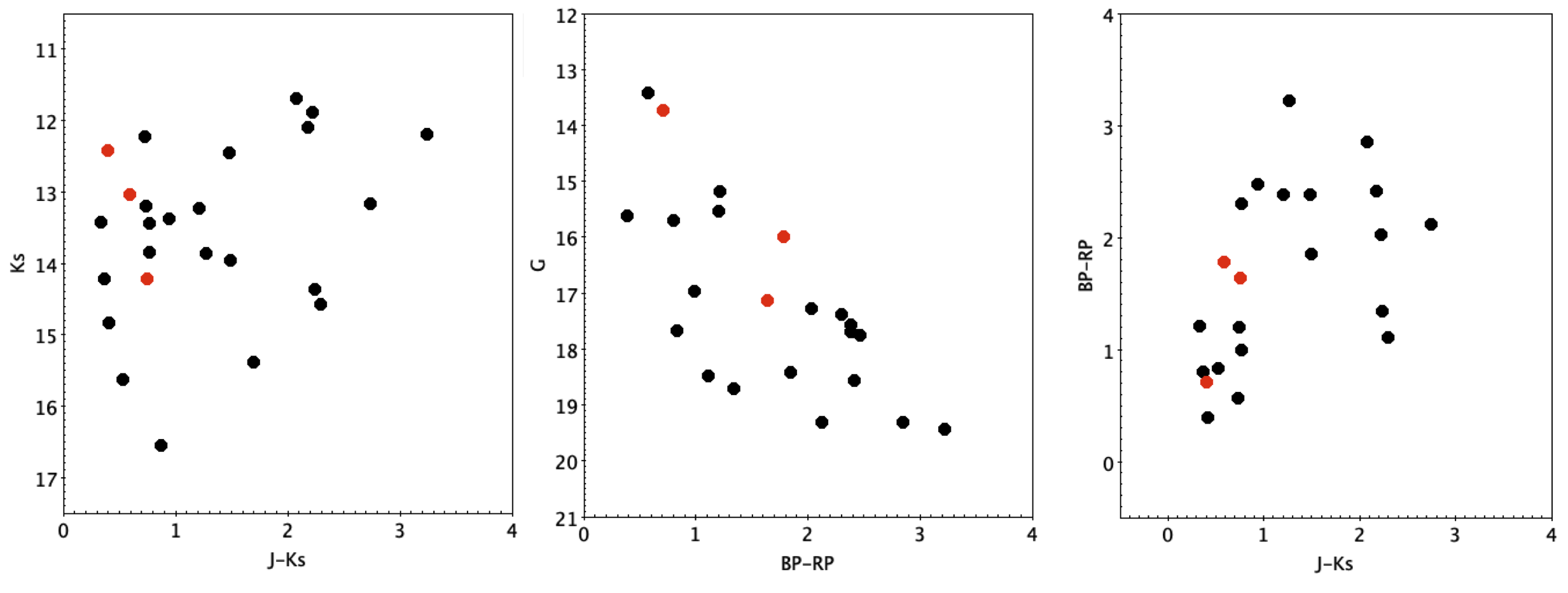}
                \caption{
 CMDs and CCDs resulting of the match between the VIVACE sample of NIR binaries from Molnar et al. (2022) and the CSPNe of Gonzalez-Santamaria et al. (2021) in the optical. The three binaries in common with Mowlavi et al. (2023) are highlighted in red.
}
                \label{Figure15}
        \end{center}
\end{figure*}

        \begin{figure*}
        \begin{center}
                \includegraphics[width=\linewidth]{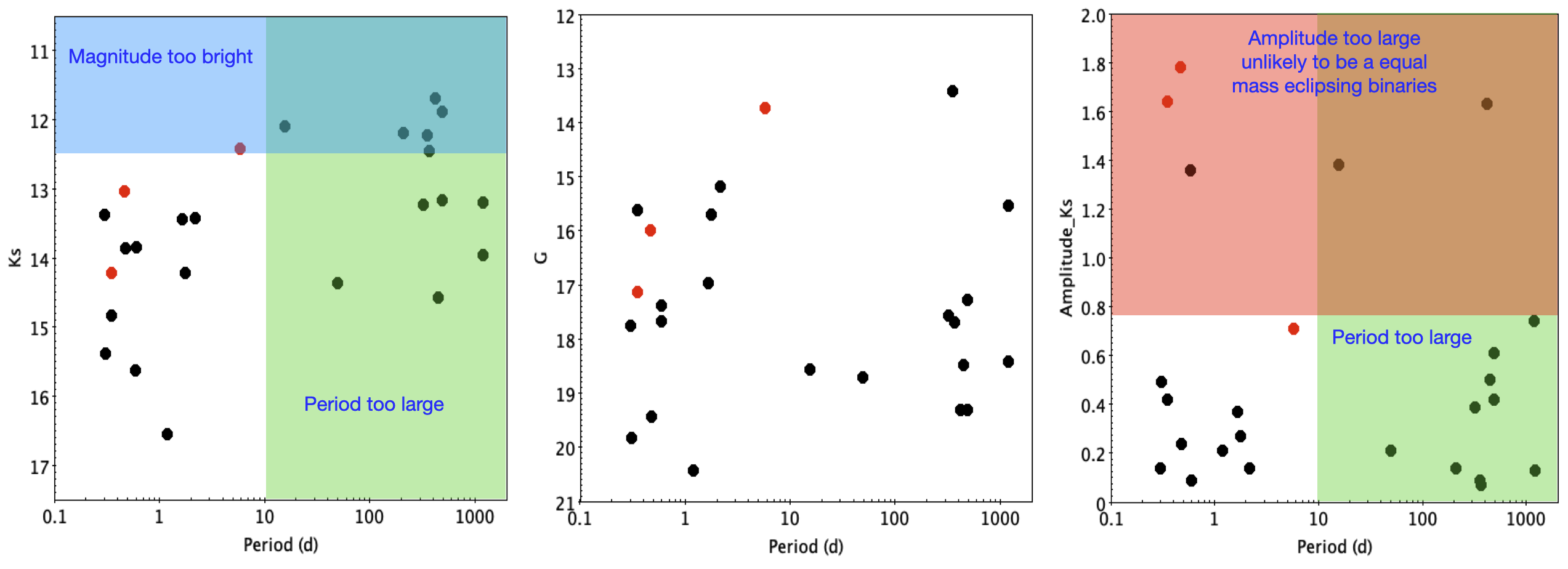}
                \caption{
Magnitude and amplitude dependence on period for the selected VIVACE binary CSPN sample, with different subsamples labeled accordingly. The three binaries in common with Mowlavi et al. (2023) are shown in red.
}
                \label{Figure16}
        \end{center}
\end{figure*}

Consequently, to uncover more binary CSPNe we inspect the sample of Gaia DR3 eclipsing binary candidates recently published by Mowlavi et al (2023).
A double match between Gonzalez-Santamaria et al. (2021) and Mowlavi et al. (2023)
with a separation $<0.7"$  reveals 
33 CSPN eclipsing binaries. 
Four of these are also identified as eclipsing binaries by Soszyński et al. (2016).
This increases the total number of CSPNe eclipsing binaries to $Ntotal = 56$ objects.

Crossmatching the two samples of eclipsing binary CSPNe (i.e., making the triple match of VIVACE + Mowlavi + Gonzalez-Santamaria)
 using a separation $<0.7$ arcsec, reveals only three CSPN eclipsing binaries in common. These three binaries in common with Mowlavi et al. (2023) are highlighted in Figures 14 and 15.
 Even though it is somewhat worrisome that not all binaries are recovered from these different variable star catalogs (possibly underscoring the incompleteness), 
 the periods measured by Molnar et al. (2022) and by Mowlavi et al. (2023) from these three different sources are all in agreement.
Nonetheless, the discovery of 56 eclipsing binary CSPNe (see Table A2), with 42 of them being the best binary candidates from our sample, can serve to confirm and complement the work of Gonzalez-Santamaria et al. (2021), cementing the idea that some of the asymmetric morphologies displayed by PNe are indeed due to binary star progenitors.
In addition, these eclipsing binary CSPNe would be prime targets for spectroscopic follow-up observations using the forthcoming  high-multiplex spectrographs like MOONS, 4MOST, and WST.

\section{Conclusions}
        \label{section9}

We have compiled a homogeneous catalog of NIR photometry for 1274 CSPNe, a valuable resource for  community studies of these objects.
Importantly, we have been able to provide NIR photometry for  $N=511$ objects in the group C sample of CSPNe from Gonzalez-Santamaria et al. (2021) that do not have Gaia colors, complementing the data for these objects.
We confirm that their NIR colors appear to be normal (following the rest of the sample), indicating that  the majority of them have been correctly identified as CSPNe.
We also explored possible misidentifications, finding only one CSPN probably misclassified as a galaxy in the literature. Moreover,  we found 56 CSPNe in total that are candidate eclipsing binary systems (making up about 2 \% of the population, but this should be considered a lower limit because of incompleteness).
We also singled out the best objects for follow-up observations.

We also investigated  objects exhibiting a large IR excess, with colors that go well beyond those attributed to their respective extinctions.
We found 14 CSPNe with large NIR excess that is indicative of the presence of cool companions and/or surrounding disks (which should also be considered a lower limit because of incompleteness).

\section{Data availability}
Table A1 of the Appendix is only available in electronic form at the CDS via anonymous ftp to cdsarc.u-strasbg.fr (130.79.128.5) or via http://cdsweb.u-strasbg.fr/cgi-bin/qcat?J/A+A/.

        \begin{acknowledgements}  
      The authors are grateful for the enlightening feedback from the referee. 
      We gratefully acknowledge the use of data from the ESO Public Survey program IDs 179.B-2002 and 198.B-2004 taken with the VISTA telescope and data products from the Cambridge Astronomical Survey Unit. 
      D.M. and M.G. are supported by Fondecyt Regular 1220724. D.M. is also supported by the BASAL Center for Astrophysics and Associated Technologies (CATA) through ANID grants ACE210002 and AFB 170002. 
      
      VF thanks Fundação Coordenação de Aperfeiçoamento de Pessoal de Nível Superior (CAPES) for granting the postdoctoral research fellowship Programa de Desenvolvimento da Pós-Graduação (PDPG)-Pós Doutorado Estratégico, Edital nº16/2022, Coordenação de Aperfeiçoamento de Pessoal de Nível Superior - Brasil (CAPES) – Finance Code 001 (88887.838401/2023-00) and Funda\c{c}\~ao de Amparo \`a Pesquisa do Estado do Rio de Janeiro (FAPERJ) for granting the postdoctoral research fellowships E-26/200.181/2025 and 200.181/2025(304980). 
      
      J.A.-G. acknowledges support from DGI-UAntof and Mineduc-UA Cod. 2355
      
      AC acknowledges the Fundação de Amparo à Pesquisa do Estado do Rio de Janeiro (FAPERJ) grant E26/202.607/2022 e 210.371/2022(270993) and  the 'Bolsa de Produtividade do CNPQ level 2'.
      
      This research has made use of the SIMBAD database, CDS, Strasbourg Astronomical Observatory, France
\end{acknowledgements}

\begin{appendix}
\section{Associated tables}

\begin{landscape}
\begin{table}[]
\caption{VVV+VVVX near-IR photometry of the studied CSPNe}
\label{tab:TabA1}
\begin{tabular}{llllcccccccl@{\hspace{2pt}}c}
\hline \hline
PN ID           & RA (deg) & DEC (dec) & G mag & BP-RP & \multicolumn{1}{c}{Z mag} & \multicolumn{1}{c}{Y mag} & \multicolumn{1}{c}{J mag} & \multicolumn{1}{c}{H mag} & \multicolumn{1}{c}{Ks mag} & Group & M\tablefootmark{a} & dang\tablefootmark{b} \\
\hline
PN G000.0+02.0  & 264.4289 & -27.8195  & 19.51 & 2.23  & 16.198 ± 0.013            & 15.929 ± 0.014            &             ...              & 14.537 ± 0.021            &             ...               & C & S &  6.9  \\
PN G000.0$-$01.3  & 267.7513 & -29.5640  & 20.14 & 0.00  & 17.506 ± 0.033            & 16.673 ± 0.027            & 15.798 ± 0.023            & 14.947 ± 0.019            & 14.475 ± 0.016             & C & E  & 12    \\
PN G000.0$-$01.8  & 268.2436 & -29.8901  & 15.85 & 2.26  & 13.989 ± 0.008            & 13.561 ± 0.006            & 12.954 ± 0.007            & 12.305 ± 0.008            & 12.068 ± 0.008             & C  & E  &  17   \\
PN G000.0$-$02.1  & 268.5179 & -29.9579  & 18.68 & 0.00  & 17.271 ± 0.025            & 17.076 ± 0.028            & 16.702 ± 0.040            & 16.299 ± 0.054            & 16.251 ± 0.073             & C  & E  &  16   \\
PN G000.0$-$02.5  & 268.9702 & -30.2613  & 19.45 & 1.91  &      ...                     &      ...                     & 16.056 ± 0.032            & 15.503 ± 0.037            & 15.206 ± 0.035             & B  & R  & 12.1 \\
PN G000.0$-$02.9  & 269.3093 & -30.3645  & 18.42 & 1.85  & 16.328 ± 0.022            & 15.933 ± 0.021            &          ...                 & 14.624 ± 0.017            &               ...             & B   & R & 38   \\
PN G000.0$-$06.8  & 273.3248 & -32.3286  & 14.36 & 0.77  & 13.776 ± 0.022            & 13.866 ± 0.026            &            ...               & 13.628 ± 0.032            &                 ...           & B   & R & 5   \\
PN G000.1+01.9  & 264.5683 & -27.8019  & 18.55 & 0.00  & 16.043 ± 0.063            & 15.188 ± 0.053            &              ...             & 13.211 ± 0.048            & 12.770 ± 0.047             & C  & R &  13   \\
PN G000.1+02.0  & 264.5131 & -27.7720  & 20.65 & 2.51  & 18.028 ± 0.039            & 17.406 ± 0.036            & 16.600 ± 0.029            & 15.801 ± 0.038            & 15.571 ± 0.042             & C  &  R &  16   \\
PN G000.1+02.6  & 263.8977 & -27.4015  & 17.94 & 2.41  & 15.708 ± 0.007            & 15.133 ± 0.007            & 14.378 ± 0.005            & 13.524 ± 0.007            & 13.217 ± 0.008             & B   & E & 10.5  \\
PN G000.1+04.3  & 262.3473 & -26.4346  & 18.75 & 1.29  & 15.474 ± 0.067            & 15.542 ± 0.113            & 14.416 ± 0.064            & 13.952 ± 0.056            & 13.277 ± 0.061             & A   & E & 5   \\
PN G000.1$-$01.0  & 267.5433 & -29.3622  & 18.47 & 0.00  & 15.704 ± 0.020            & 14.618 ± 0.016            & 13.524 ± 0.020            &        ...                   & 11.854 ± 0.011             & C  & E & 14    \\
PN G000.1$-$01.7  & 268.2034 & -29.6994  & 15.97 & 2.79  & 13.873 ± 0.006            & 13.306 ± 0.006            & 12.558 ± 0.007            & 11.682 ± 0.006            & 11.415 ± 0.010             & C  & E & 16.7    \\
PN G000.1$-$01.9  & 268.3492 & -29.8300  & 17.64 & 2.33  & 15.601 ± 0.011            & 15.025 ± 0.013            &           ...                & 13.546 ± 0.014            & 13.212 ± 0.013             & B  & E &  14   \\
PN G000.1$-$05.6  & 272.1281 & -31.6098  & 18.46 & 1.44  & 17.037 ± 0.013            & 16.663 ± 0.016            & 16.162 ± 0.013            & 15.580 ± 0.018            & 15.389 ± 0.026             & B  & E &  18.3   \\
PN G000.2+01.4  & 265.1607 & -28.0172  & 21.06 & 0.00  & 17.522 ± 0.040            & 17.096 ± 0.043            & 16.380 ± 0.093            & 15.686 ± 0.061            & 14.176 ± 0.058             & C  & E &  6.5   \\
PN G000.2+06.1  & 260.7228 & -25.4169  & 20.27 &  ...  &              ...             &           ...                & 17.670                    & 16.951                    & 16.774                     & C   & B & 16   \\
PN G000.2$-$01.9  & 268.4402 & -29.7297  & 16.88 & 1.34  & 15.595 ± 0.021            &     ...                      & 15.411 ± 0.028            & 15.284 ± 0.058            & 15.245 ± 0.106             & B   & B & 20   \\
PN G000.2$-$01.9a & 268.4195 & -29.7107  & 16.67 & 1.47  & 15.414 ± 0.010            & 15.233 ± 0.011            & 14.812 ± 0.014            & 14.338 ± 0.020            & 14.259 ± 0.020             & B   & E & 50   \\
PN G000.2$-$03.4  & 269.9484 & -30.5094  & 19.20 & 1.33  & 17.705 ± 0.033            & 17.367 ± 0.035            & 16.867 ± 0.030            & 16.330 ± 0.043            & 16.251 ± 0.063             & C   & E & 19   \\
PN G000.2$-$03.4a & 269.9997 & -30.4352  & 20.42 & 0.00  & 18.112 ± 0.046            & 17.789 ± 0.052            &         ...                  & 16.893 ± 0.088            & 16.639 ± 0.120             & C  & S &  48   \\
PN G000.2$-$04.6  & 271.1840 & -31.0469  & 17.84 & 1.79  & 16.038 ± 0.011            & 15.590 ± 0.012            & 14.976 ± 0.009            & 14.325 ± 0.008            & 14.141 ± 0.014             & C   & E &  6.4  \\
PN G000.3+01.5  & 265.0960 & -27.8200  & 21.09 & 0.00  & 17.795 ± 0.031            & 18.030 ± 0.071            & 16.474 ± 0.034            & 15.770 ± 0.030            & 14.370 ± 0.021             & C  & S & 3  \\
PN G000.3+01.7  & 264.9672 & -27.7386  & 20.65 & 0.00  & 18.527 ± 0.057            & 17.928 ± 0.052            & 17.117 ± 0.044            & 16.406 ± 0.042            & 16.139 ± 0.065             & C  & E &  21   \\
PN G000.3+03.2  & 263.4691 & -26.9242  & 20.36 & 1.23  & 19.001 ± 0.124            & 18.930 ± 0.178            & 18.447 ± 0.116            &          ...                 &            ...                & A  & E &  9   \\
PN G000.3+03.4  & 263.3638 & -26.8000  & 20.41 & 0.00  & 18.358 ± 0.059            & 17.956 ± 0.072            & 17.283 ± 0.044            & 16.691 ± 0.047            & 16.532 ± 0.090             & C  & R &  15   \\
PN G000.3+04.2  & 262.5515 & -26.3504  & 18.47 & 2.03  & 16.549 ± 0.012            & 16.196 ± 0.012            & 15.820 ± 0.011            & 15.332 ± 0.014            & 15.057 ± 0.022             & C   & R & 5   \\
PN G000.3+04.5  & 262.2806 & -26.2455  & 20.69 & 2.13  & 18.660 ± 0.069            & 18.040 ± 0.068            & 17.461 ± 0.042            & 16.771 ± 0.052            & 16.413 ± 0.076             & B  & B &  17   \\
PN G000.3+06.9  & 260.0907 & -24.8639  & 20.62 & 1.47  &             ...              &             ...              & 18.302                    & 17.810                    & 17.349                     & C  & E &  15.2   \\
PN G000.3+07.3  & 259.6779 & -24.6898  & 20.47 & 1.02  &            ...               &            ...               & 18.762                    & 18.266                    &               ...             & B   & E &  23  \\
PN G000.3$-$02.8  & 269.4307 & -30.0421  & 19.13 & 1.37  & 17.552 ± 0.036            & 17.181 ± 0.029            & 16.746 ± 0.029            & 16.257 ± 0.037            & 16.080 ± 0.052             & C  & R &  9   \\
PN G000.3$-$03.4  & 270.0462 & -30.3971  & 18.12 & 0.89  & 17.262 ± 0.024            & 17.173 ± 0.029            & 16.963 ± 0.035            & 16.807 ± 0.068            & 16.596 ± 0.096             & C  & R &  5   \\
PN G000.3$-$04.2  & 270.8405 & -30.7265  & 15.38 & 2.65  & 13.437 ± 0.008            & 12.874 ± 0.008            & 12.159 ± 0.004            &       ...                    &           ...                 & B   & R &  36  \\
PN G000.3$-$04.2a & 270.8575 & -30.7735  & 20.30 & 0.00  & 18.779 ± 0.052            & 18.478 ± 0.068            & 18.057 ± 0.076            & 17.384 ± 0.130            &            ...                & C   & E & 18   \\
PN G000.3$-$04.6  & 271.2611 & -30.9713  & 19.63 & 1.06  & 17.389 ± 0.053            & 17.225 ± 0.061            & 16.645 ± 0.042            & 16.456 ± 0.057            & 16.094 ± 0.077             & B  & E &  5.5   \\
PN G000.4+02.2  & 264.5310 & -27.3375  & 17.52 & 2.34  & 14.810 ± 0.008            & 14.389 ± 0.007            & 13.594 ± 0.010            & 13.114 ± 0.010            & 12.415 ± 0.012             & C   & S & 6   \\
PN G000.4+04.4  & 262.4682 & -26.1871  & 19.18 & 1.76  & 16.993 ± 0.022            & 16.937 ± 0.031            & 16.439 ± 0.028            & 16.025 ± 0.027            & 15.721 ± 0.046             & B  & R &  9   \\
PN G000.4$-$01.3  & 268.0159 & -29.2780  & 20.55 & 0.00  & 18.353 ± 0.053            & 17.899 ± 0.062            & 17.178 ± 0.061            & 16.512 ± 0.070            & 16.393 ± 0.107             & C   & E &  31.5  \\
PN G000.4$-$02.1  & 268.8577 & -29.6418  & 18.82 & 0.00  & 17.364 ± 0.053            &        ...                   & 16.528 ± 0.052            & 15.915 ± 0.058            & 15.950 ± 0.069             & C   &  ... &  33  \\
PN G000.4$-$02.9  & 269.5808 & -30.0109  & 18.34 & 0.82  & 17.159 ± 0.028            & 17.133 ± 0.046            & 17.102 ± 0.069            &        ...                   &           ...                 & A  & R &  7.2  \\
\hline
\end{tabular}

\tablefoot{
VVV+VVVX near-IR photometry of a sample of 40 CSPNe identified by Gonzalez-Santamaria et al. (2021). The full catalog is available at the CDS.\\

\tablefoottext{a}{PN morphology from Parker etl al. (2016) as E: Elliptical, R: Round, B: Bipolar, I: Irregular, A: Asymmetric, S: quasi-Stellar.}  

\tablefoottext{b}{PNe major diameters in arcsec from Parker et al. (2016).}  
}

\end{table}

\end{landscape}

\begin{table*}[]
\centering
\caption{Eclipsing binary CSPNe.}
\label{tab:TabA2}
\begin{tabular}{lllccccccc}
\hline \hline
PN ID & RA (deg) & DEC (deg) & G mag  & BP-RP & J mag & Ks mag & Ks amp & P (days) & Group \\
\hline
\multicolumn{10}{l}{Part I. CSPNe identified as binaries by both Molnar et al. (2022) and  Mowlavi et al (2023)}  \\
\hline
PN G314.6$-$00.1 & 217.4712 & -60.7323  & 17.13 & 1.64  & 14.97             & 14.22              & 0.38               & 0.35          & A     \\
PN G329.0+01.9 & 237.9206 & -51.5246  & 13.73 & 0.71  & 12.82             & 12.42              & 0.14               & 5.81          & A     \\
PN G001.0+01.4 & 265.6169 & -27.2255  & 15.99 & 1.78  & 13.62             & 13.03              & 0.32               & 0.47          & B     \\
\hline
\multicolumn{9}{l}{Part II. CSPNe identified as binaries by Molnar et al. (2022)}                                 \\
\hline
PN G305.9$-$01.2 & 199.7114 & -63.9818  & 19.32 & 2.85  & 13.77             & 11.69              & 1.63               & 418.79        & B     \\
PN G308.4$-$00.1 & 204.9849 & -62.5091  & 19.32 & 2.12  & 15.91             & 13.17              & 0.42               & 486.17        & C     \\
PN G321.0+01.4 & 227.1294 & -56.5504  & 19.43 & 3.22  & 15.13             & 13.86              & 0.24               & 0.48          & B     \\
PNG 330.5$-$01.4 & 243.4235 & -53.0809  & 21.12 & ...  & 15.44             & 12.20              & 0.14               & 208.66        & C     \\
PN G344.4+01.8 & 253.6799 & -40.6962  & 19.82 &  ...  & 17.08             & 15.38              & 0.49               & 0.31          & C     \\
PN G349.3$-$01.1 & 260.5653 & -38.4838  & 15.62 & 0.39  & 15.25             & 14.84              & 0.42               & 0.35          & A     \\
PN G357.5+03.1 & 261.8515 & -29.3541  & 15.18 & 1.21  & 13.76             & 13.43              & 0.14               & 2.17          & B     \\
PNG 359.9+03.7 & 262.7367 & -26.9864  & 17.56 & 2.38  & 14.44             & 13.23              & 0.39               & 322.05        & B     \\
PN G353.6$-$01.3 & 263.8199 & -35.0472  & 17.76 & 2.47  & 14.32             & 13.38              & 0.14               & 0.30          & A     \\
PN G003.5+04.3 & 264.3685 & -23.6920  & 18.57 & 2.41  & 14.28             & 12.10              & 1.38               & 15.58         & C     \\
PN G004.1+03.6 & 265.3701 & -23.5387  & 18.49 & 1.11  & 16.87             & 14.58              & 0.50               & 448.13        & B     \\
PN G355.2$-$02.5 & 266.0578 & -34.2925  & 16.97 & 0.99  & 14.21             & 13.44              & 0.37               & 1.66          & C     \\
PN G355.5$-$02.8 & 266.5770 & -34.2103  & 15.54 & 1.20  & 13.93             & 13.19              & 0.74               & 1199.33       & B     \\
PN G354.5$-$03.9 & 267.0681 & -35.6418  & 17.68 & 0.83  & 16.16             & 15.63              & 1.36               & 0.59          & B     \\
PN G004.1+01.7 & 267.1344 & -24.4526  & 17.28 & 2.03  & 14.11             & 11.89              & 0.61               & 486.29        & B     \\
PN G358.9$-$01.5 & 267.3334 & -30.6017  & 17.69 & 2.38  & 13.94             & 12.46              & 0.07               & 365.17        & B     \\
PN G007.3+01.7 & 268.8948 & -21.7104  & 17.39 & 2.30  & 14.62             & 13.85              & 0.09               & 0.60          & B     \\
PN G357.9$-$03.8 & 269.0581 & -32.6228  & 18.70 & 1.34  & 16.60             & 14.36              & 0.21               & 49.43         & B     \\
PN G000.0$-$02.9 & 269.3093 & -30.3645  & 18.42 & 1.85  & 15.45             & 13.96              & 0.13               & 1205.40       & B     \\
PN G005.3$-$02.0 & 271.3671 & -25.3407  & 20.43 &  ...   & 17.42             & 16.55              & 0.21               & 1.19          & C     \\
PN G002.2$-$06.3 & 274.0806 & -30.1267  & 13.43 & 0.57  & 12.96             & 12.23              & 0.09               & 354.74        & B     \\
PN G009.7$-$03.9 & 275.5353 & -22.3854  & 15.70 & 0.80  & 14.59             & 14.22              & 0.27               & 1.76          & B     \\
\hline
\multicolumn{9}{l}{Part III. CSPNe identified as binaries by Mowlavi et al (2023)}                                 \\
\hline
PN G164.8$-$09.8 & 66.3619  & 35.1022   & 18.21 & 1.75  &       ...            &         ...           &        ...            & 0.54          & B     \\
PN G221.3$-$12.3 & 95.4283  & -12.9872  & 17.50 & -0.16 &       ...            &        ...            &          ...          & 0.21          & A     \\
PN G240.3$-$07.6 & 108.7080 & -27.8398  & 16.26 & 0.49  &       ...            &        ...            &       ...             & 1.88          & C     \\
PN G259.1+00.9 & 129.2848 & -39.4190  & 17.54 & 1.82  &      ...            &        ...            &       ...        & 1.22          & A     \\
PN G268.9$-$00.4 & 136.4210 & -47.9014  & 12.14 & 1.54  &            ...            &        ...            &       ...        & 0.61          & A     \\
PN G283.7$-$05.1 & 149.6339 & -61.4452  & 17.62 & 0.05  &         ...            &        ...            &       ...        & 0.49          & B     \\
PN G283.6+25.3 & 171.6824 & -34.3697  & 16.68 & 0.39  &             ...            &        ...            &       ...          & 11.51         & B     \\
PN G135.9+55.9 & 178.3531 & 59.6658   & 18.03 & -0.29 &          ...            &        ...            &       ...          & 0.33          & A     \\
PN G311.0+02.4 & 208.9301 & -59.3777  & 11.89 & 0.65  &          ...            &        ...            &       ...           & 4.91          & A     \\
PN G337.0+08.4 & 240.5756 & -41.4471  & 16.73 & 0.87  &             ...            &        ...            &       ...        & 0.59          & B     \\
PN G000.1+17.2 & 250.9741 & -18.9533  & 15.17 & 0.62  &          ...            &        ...            &       ...             & 0.59          & B     \\
PN G335.2$-$03.6 & 251.2508 & -51.2059  & 16.78 & 0.73  &          ...            &        ...            &       ...           & 3.48          & A     \\
PN G359.8+05.2 & 261.3484 & -26.1981  & 19.04 & 1.14  &           ...            &        ...            &       ...          & 0.45          & B     \\
PN G009.6+10.5 & 262.2584 & -15.2179  & 16.24 & 0.27  &           ...            &        ...            &       ...           & 0.23          & A     \\
PN G003.1+05.2 & 263.2793 & -23.4668  & 19.30 & ...  &           ...            &        ...            &       ...          & 25.11         & C     \\
PN G355.3$-$03.2 & 266.7847 & -34.5953  & 18.78 & 0.85  &            ...            &        ...            &       ...           & 0.45          & A     \\
PN G006.3+03.3 & 266.8914 & -21.7898  & 18.47 & 1.57  &        ...            &        ...            &       ...         & 0.25          & B     \\
PN G000.6$-$01.3 & 268.1502 & -29.1110  & 19.24 & 2.03  &       ...            &        ...            &       ...           & 0.72          & B     \\
PN G359.1$-$02.3 & 268.1919 & -30.8264  & 17.07 & 1.11  &            ...            &        ...            &       ...          & 1.61          & A     \\
PN G001.5$-$02.8 & 270.0934 & -29.0776  & 18.15 & ...   &       ...            &        ...            &       ...          & 0.47          & C     \\
PN G009.1$-$02.7 & 274.0824 & -22.3804  & 18.77 & 1.30  &       ...            &        ...            &       ...        & 0.45          & B     \\
PN G004.2$-$05.9 & 274.6598 & -28.1330  & 17.32 & 0.44  &           ...            &        ...            &       ...          & 1.22          & B     \\
PN G357.2$-$09.8 & 275.0385 & -36.1226  & 16.94 & 0.21  &          ...            &        ...            &       ...            & 0.68          & B     \\
PN G024.1+01.1 & 277.8198 & -7.2556   & 19.82 & 2.37  &          ...            &        ...            &       ...          & 0.62          & C     \\
PN G055.4+16.0 & 277.8275 & 26.9367   & 14.96 & -0.23 &      ...            &        ...            &       ...        & 0.94          & A     \\
PN G005.1$-$08.9 & 278.1287 & -28.7223  & 17.16 & -0.01 &        ...            &        ...            &       ...           & 0.80          & A     \\
PN G054.5+01.8 & 291.3954 & 20.0596   & 18.64 & 1.38  &       ...            &        ...            &       ...          & 0.85          & B     \\
PN G053.8$-$03.0 & 295.5429 & 17.0873   & 14.97 & 0.28  &          ...            &        ...            &       ...            & 0.93          & B     \\
PN G061.5$-$02.6 & 299.3468 & 23.8801   & 18.10 & 1.28  &       ...            &        ...            &       ...        & 2.34          & A     \\
PN G056.8$-$06.9 & 300.6514 & 17.6141   & 17.23 & -0.06 &       ...            &        ...            &       ...          & 0.56          & A  \\
\hline
\end{tabular}
\tablefoot{
CSPNe from Gonzalez-Santamaria et al. (2021) identified as eclipsing binaries by Molnar et al. (2022) and  Mowlavi et al (2023).}

\end{table*}

\end{appendix}

\end{document}